\documentclass[11pt,a4paper]{article}
\usepackage[utf8]{inputenc}
\usepackage[english]{babel}
\usepackage{amsmath}
\usepackage{amsfonts}
\usepackage{amssymb}
\usepackage{amsthm}
\usepackage{color}
\usepackage{graphicx}
\usepackage[force]{feynmp-auto}
\usepackage{cite}
\usepackage{url,hyperref}

\def\T{\tilde{T}}

 \def\be{\begin{equation}}
\def\ee{\end{equation}}
 \def\ba{\begin{align}}
\def\ea{\end{align}}
\def\bea{\begin{eqnarray}}
\def\eea{\end{eqnarray}}
\newcommand{\bseq}{\begin{subequations}}
\newcommand{\eseq}{\end{subequations}}

\def\a{\alpha}
\def\b{\beta}

\def\m{\mu}
\def\n{\nu}

\begin{document}
\title{
\vspace{0.5cm} 
{\bf What do gravitons say about (unimodular) gravity?}}

\author{Mario Herrero-Valea\\[2mm]
{\normalsize\it Institute of Physics, Laboratory of Particle Physics and Cosmology}\\[-1mm]
{\normalsize\it  Ecole Polytechnique F\'ed\'erale de Lausanne,}\\[-1mm]
{\normalsize\it CH-1015, Lausanne, Switzerland}\\}
\date{}
\maketitle
\begin{abstract}
\vspace*{-.4cm}
We revisit the problem of constraining the weak field limit of the gravitational lagrangian from S-matrix properties. From unitarity and Lorentz invariance of the S-matrix of massless gravitons, we derive on-shell gauge invariance to consist on the transverse part of the linearised diffeomorphisms group. Moreover, by looking to the interaction between sources, we conclude that there exist only two possible lagrangians that lead to a well-defined covariant interaction, corresponding to the weak field limits of General Relativity and Unimodular Gravity. Additionally, this result confirms the equivalence of the S-matrix of both theories around flat space-time.
\end{abstract}

\section{Introduction}
Perturbative Quantum Field Theory (QFT) has proven to be one of the most powerful tools in making predictions about physical quantities. Starting from a lagrangian and by applying a finite set of rules, we are able to compute observable quantities with a great precision (at least in the perturbative regime). In particular, the formalism allows, via the Quantum Effective Action and the LSZ formula, to obtain the expression for the scattering matrix (S-matrix) of any given theory, an object which presumably contains all possible information on physical observables of an interacting theory about a flat space-time. However, there is a big conceptual gap between the Lagrangian and the S-matrix, and one could question if there exist a single choice of Lagrangian for every given S-matrix. In other words, how many different lagrangians could lead to the same physical predictions or, rephrasing it in yet another way, what are the minimal requirements for a given lagrangian to reproduce what we know from experiments and basic assumptions. This is of particular relevance in fields where our available data is limited or even meagre, as it is the case for Beyond Standard Model (BSM) physics or for the quantum regime of gravitational interactions. The lack of data in these fields have led to a fruitful development of multiple alternative theories and effective field theories, which aim to explain a very reduced number of observations, whenever these exist, and to produce a new set of predictions that could be contrasted in future experiments, altogether with possible solutions to open theoretical problems.

One of these everlasting problems is \emph{the cosmological constant problem}\cite{Weinberg:1988cp, Martin:2012bt}. In particular, the fact that radiative corrections to vacuum energy, even those coming from Standard Model or BSM particles in absence of Quantum Gravity, are brobdingnagian when compared to the experimental value derived from observations. Although in the presence of a bare cosmological constant in the gravitational lagrangian this is just a fine tuning problem, the fact that we need to adjust the value of the cosmological constant $\Lambda$ with a precision of $120$ orders of magnitude, makes this an unpleasant hierarchy problem much worse than the one of the Higgs boson mass. Likewise the latter, the existence of supersymmetry (SUSY) could relax the numerical clash, but the absence of any clue of the existence of SUSY in LHC data puts this solution in tension with naturalness. 

Another possible solution that has attracted attention in the recent years although it is almost as old as GR itself\footnote{The equations of motion of Unimodular Gravity were actually proposed by Einstein himself in 1919\cite{Einstein}, although his goal was not to solve the cosmological constant problem but to try to explain the structure of the electron.} is provided by modifying the infra-red limit of gravitational interactions in Unimodular Gravity (UG)\cite{Alvarez:2005iy,Unruh:1988in,Henneaux:1989zc,Kreuzer:1989ec,Ng:1999tx, Buchmuller:1988wx}. The core idea of UG relies on the fact that the Einstein equations of GR are not independent, but only their traceless part is dynamically non-trivial. Bianchi identities must be always satisfied though, and they provide an extra constraint in order to recover the trace part and the whole set of equations. The cosmological constant, being contained only in the trace part, decouples and it is substituted in the solutions to the equations of motion by an integration constant unrelated to the vacuum energy at all. Thus, fixing its value is a matter of fixing initial and boundary conditions.

From a lagrangian point of view, UG does not differ much from GR\cite{Gielen:2018pvk}. It is obtained from the Einstein-Hilbert action, which must be however varied under the condition of unit determinant for the metric tensor $|g|=1$ and thus, as a requirement for this constraint to hold, the theory is only invariant under volume preserving diffeomorphisms. Additionally, since $\sqrt{|g|}=1$, any coupling to the cosmological constant, or to any vacuum energy and its radiative corrections, is dropped out of the dynamics of the theory at any loop order. The effective cosmological constant of the theory is, again, an integration constant obtained when solving the equations of motion for the mean field. In the weak field limit, the dynamics of the theory is described by an action with a gauge symmetry being the product of transverse infinitesimal diffeomorphims and local rescalings\footnote{These act by shifting the value of the graviton field $h_{\m\n}$ by a quantity proportional to its trace $h_{\m\n}\rightarrow h_{\m\n}+\phi \eta_{\m\n}$ with $\phi$ an arbitrary function.}, a group that has been dubbed \emph{WTDiff} in previous works. 

In the recent years there have been various interesting advances in understanding UG and the solidness of its proposal as alternative to GR. These extend from exhaustive studies of how to embed cosmological models into it\cite{Ellis:2010uc,Alvarez:2007nn,Nozari:2017ocg, Shaposhnikov:2008xb} to a thorough research of the ultra-violet regime of the theory\cite{Alvarez:2015pla,Alvarez:2015sba, Eichhorn:2015bna, Saltas:2014cta,Smolin:2009ti}; passing by several consistency studies of its interaction with matter fields\cite{Alvarez:2016uog,Martin:2017ewb,Gonzalez-Martin:2017bvw, Ardon:2017atk,Gonzalez-Martin:2017fwz, Gonzalez-Martin:2018dmy, Percacci:2017fsy, Carballo-Rubio:2015kaa}. Of particular interest is the work of \cite{Barcelo:2014mua}, where it is proven that a \emph{WTDiff} invariant theory is the only other possible Lorentz invariant theory with a massless spin two field, besides the weak limit of GR, that can be completed to a non-linear theory by following Deser's recipe\cite{Padmanabhan:2004xk,Deser:2009fq,Deser:1969wk,Deser:2017ihr}. Of course, the completion they find is Unimodular Gravity. 

Motivated by the fact that Unimodular Gravity seems to be as good as General Relativity for classical matters, while it seems to provide a solution to the cosmological constant problem, we wish to investigate this dichotomy from 
the viewpoint of massless gravitons and their S-matrix. Explicitly, we wish to revisit the standard lore that relativistic massless gravitons lead uniquely to the weak field limit of GR\cite{Weinberg:1964ew,Weinberg:1965rz}, a fact that cannot be true in full generality precisely because of the existence of UG. Thus, we will study what are the minimal requirements to build a theory of massless gravitons and where the bifurcation appears. Additionally, we will answer two long standing questions about UG that are not well-understood in the literature yet -- how does the equivalence principle arises in the theory\cite{Ortin:2017fux} and what is the relation of its S-matrix with the one of GR.

The content of this paper is organized as follows. In section 2 we introduce several basic notions about physical states of gravitons and we derive the transformation rules for the polarizations. Later on, in section 3 we derive the form of the on-shell gauge invariance preserved by the physical states by requiring unitarity and Lorentz invariance of the S-matrix. In section 4 we use this to derive a weak field lagrangian which leads to a covariant interaction and we show how this lagrangian corresponds to the gauge fixed limit of known theories in section 5. Finally, we devote section 6 to discuss about the non-linear completion of the theory for gravitons and we draw conclusions in section 7.

\section{Integer spin particles}
We will closely follow \cite{Weinberg:1964ew,Weinberg:1965rz} in what follows. Although we are interested in gravity and thus in gravitons, it will be useful to consider also photons, since their construction is similar and they serve as a good simpler example of what we are studying. 

We introduce photons and gravitons as massless particles with integer spin that move along a light-like trajectory in flat space-time with momentum $q^\m$ satisfying\footnote{Throughout this work will use the mostly-plus convention for the signature of the flat metric $\eta_{\m\n}=\text{Diag}(-1,1,1,1)$. Greek indices run in the range $\{0,1,2,3\}$, while latin indices run only over space directions.}
\begin{align}\label{eq:on-shell}
q^2 = q^\m q_\m=0
\end{align}

The physical polarizations of a given particle with spin $s$ are objects $\epsilon^{\m_1\m_2...\m_s}$ that transform under the irreducible representations of the Lorentz group that preserve the on-shell condition \eqref{eq:on-shell}. This is the little group $\mathbb{L}$, whose elements $L^{\m}_{\,\,\,\n}$ thus satisfy
\begin{align}
L^\m_{\,\,\,\n}q^\m=q^\m
\end{align}
where $q^\m$ is any light-like vector. In particular, to give a clear characterization of $\mathbb{L}$, we can take a particular momentum $K^\mu$ in the $(t,x^3)$ plane
\begin{align}
K^\m=(\kappa,0,0,\kappa)
\end{align}
with $\kappa$ arbitrary.

With this given vector, the little group corresponds to all those transformations that take the euclidean plane $(x^1,x^2)$ onto itself. That is, we have $\mathbb{L}=E^2$. Any element of this group will consist of a translation plus a rotation, with the latter being represented by a unitary operator
\begin{align}\label{eq:rotation}
U[L]=e^{iJ_3 \theta[L]}
\end{align}
where $J_3$ is the helicity of the particle along the $x^3$ direction, which for massless particles is aligned with their spin, and $\theta[L]$ is the rotation angle, different for every element of the little group $\mathbb{L}$. 

For concreteness, any element of the little group can be described by the rotation angle $\theta[L]$ and a vector $X=(0,X^1,X^2,0)$ through the matrix
\begin{align}\label{eq:lg_matrix}
{\cal E}=\begin{pmatrix}
\cos \theta[L] & \sin \theta[L] & 0 &0\\
-\sin \theta[L] & \cos \theta[L] & 0 &0\\
0&0&1&0\\
0&0&0&1
\end{pmatrix}\begin{pmatrix}
1&0&-X^1 & X^1\\
0&1&-X^2&X^2\\
X^1&X^2&1-\frac{(X^1)^2+(X^2)^2}{2}&\frac{(X^1)^2+(X^2)^2}{2}\\
X^1&X^2&-\frac{(X^1)^2+(X^2)^2}{2}&1+\frac{(X^1)^2+(X^2)^2}{2}
\end{pmatrix}
\end{align}

The corresponding polarizations can now be constructed easily in the frame where $q^\mu=K^\m$ starting from the corresponding ones for a spin one particle. There, the orthogonal euclidean plane can be parametrized by using the following basis
\begin{align}
e^\mu_{\pm}=\frac{1}{\sqrt{2}}(0,1,\pm i,0)
\end{align}
which can be directly taken to be the polarizations of a spin one particle (photon herein after) in this frame. For a different momentum $q^\mu$ we just need to act on $e^\m_\pm$ with the rotation $R^\m_{\;\;\;\n}(q)$ which aligns the $x^3$-axis with $q^\mu$
\begin{align}\label{eq:rotationR}
\epsilon^\mu_{\pm}(q)=R^\m_{\;\;\;\n}(q)e^\n_\pm
\end{align}

From this construction, it is easy to see that the photon polarization satisfies the following properties
\begin{align}\label{eq:photon_properties}
\nonumber &(\epsilon^\mu_\pm(q))^* \epsilon_{\pm\mu}(q)=1\\
\nonumber&\epsilon^\mu_\pm(q) \epsilon_{\mu\pm}(q)=0\\
\nonumber&(\epsilon^\mu_\pm(q))^* =\epsilon^\m_\mp(q)\\
&\epsilon^0_{\pm}(q)=q_\m \epsilon^\m_\pm=0
\end{align}

Polarizations for higher spin particles belong to irreducible representations of the direct product of $\epsilon^\m_\pm(q)$. For the particular case of a spin two particle (graviton herein after), the two polarizations can be taken to be
\begin{align}\label{eq:epsilon_graviton}
\epsilon_{\pm}^{\m\n}(q)=\epsilon^\m_{\pm}(q)\epsilon^\n_{\pm}(q)
\end{align}
and they satisfy a set of properties inherited from \eqref{eq:photon_properties}. The ones that will be relevant for the rest of this work are
\begin{align}
\nonumber&\eta_{\m\n}\epsilon_{\pm}^{\m\n}(q)=0\\
\nonumber&q_\m\epsilon_{\pm}^{\m\n}(q)=0\\
\nonumber&\epsilon_{\pm}^{0\n}(q)=0\\
&\epsilon_{\pm}^{\m\n}(q)=\epsilon_{\pm}^{\n\m}(q)
\end{align}

Note that altogether these properties imply that the polarizations can actually be written by using purely space indices as $\epsilon^{ij}_\pm(q)$ and that they are both traceless and transverse. In this section we will keep the full space-time greek indices thought, but we will resort to space indices later on this work.

Although it carries a Lorentz index, the photon polarization $\epsilon^\m_\pm(q)$ is \emph{not} a vector, since it does not transform homogeneously when acting with a Lorentz transformation upon it. In other words
\begin{align}
\alpha^\m_{\,\,\,\n}\epsilon^{\n}_{\pm}(q)\neq \epsilon^\m_\pm(\alpha q)
\end{align}
where $\alpha^\m_{\,\,\,\n}$ is an arbitrary Lorentz transformation.

In order to find what is the real transformation rule, let us go to the frame where we take the momentum to be $K^\m$ by means of a Lorentz transformation defined as
\begin{align}\label{eq:def_Lambda}
q^\m=\Lambda^\m_{\;\;\;\n}(q)K^\n
\end{align}

Here $\Lambda^\m_{\;\;\;\n}(q)$ is the result of combining the rotation $R^\m_{\;\;\;\n}(q)$ in \eqref{eq:rotationR} with a boost along the $x^3$-axis which takes the vector to the appropriate length $\Lambda^\m_{\;\;\;\n}(q)=R^\m_{\;\;\;\a}(q)B^\a_{\;\;\;\n}(q)$. By using it, we rewrite the former inequality as
\begin{align}
\alpha^\m_{\,\,\,\n}\Lambda^\nu_{\,\,\,\rho}(q)e^{\rho}_{\pm}(q)\neq \Lambda^{\m}_{\,\,\,\n}(\alpha q)e^{\n}_{\pm}
\end{align} 
where we must note that the boost leaves any polarization unaffected, since they are blind to the length of the momentum.

Multiplying by the inverse of the operator in the l.h.s,
\begin{align}
e^{\m}_{\pm}\neq[\Lambda^{-1}(q)]^\m_{\,\,\,\n}[\alpha^{-1}]^\n_{\,\,\,\rho}[\Lambda(\alpha q)]^\rho_{\,\,\,\sigma}e^\sigma_{\pm}
\end{align}

Therefore, in order to study how $\epsilon^\m_\pm(q)$ behaves under a Lorentz transformation, it is enough to find what is the action of the operator in the r.h.s. of the former expression
\begin{align}
\Xi^\m_{\,\,\,\sigma}=[\Lambda^{-1}(q)]^\m_{\,\,\,\n}[\alpha^{-1}]^\n_{\,\,\,\rho}[\Lambda(\alpha q)]^\rho_{\,\,\,\sigma}
\end{align}

By taking into account the definition \eqref{eq:def_Lambda}, we see that $\Xi^\m_{\,\,\,\n}$ is an element of the little group $\mathbb{L}$, satisfying $\Xi^\m_{\,\,\,\n}K^\n=K^\m$. Its action over  $e^\m_\pm$ it is obtained from \eqref{eq:lg_matrix} and can be written as a rotation, as given in \eqref{eq:rotation}, plus a translation along $K^\m$
\begin{align}
\Xi^\m_{\,\,\,\n}e^\n_\pm=e^{\pm i \theta[\Xi]}e^\m_\pm+X_\pm K^\m
\end{align}
where $X_\pm$ are the components of the translation and $J_3=1$ for a spin one massless particle.

Acting with $\Xi^\m_{\,\,\,\n}$ on $e^\m_\pm$ and applying $\Lambda^\m_{\,\,\,\n}(q)$ on both sides, this can be rewritten as
\begin{align}
[\alpha^{-1}]^\m_{\,\,\,\n}\epsilon^\n_\pm (\alpha q)=e^{\pm i \theta[\Xi]}\epsilon^\m_\pm(q)+X_\pm q^\m
\end{align}

The form of $X_\pm$ can be obtained by setting $\mu=0$ in this expression, which allows us to rewrite it as 
\begin{align}\label{eq:Lorentz_epsilon}
\left([\alpha^{-1}]^\m_{\,\,\,\n}-[\alpha^{-1}]^0_{\,\,\,\n}\frac{q^\m}{|q|}\right)\epsilon^{\n}_{\pm}(\alpha q)=e^{\pm i \theta[\Xi]}\epsilon^\m_\pm(q)
\end{align}
where $|q|$ is the length of the space component of the momentum $q^i$.

Note that if it where not because of the second term on the l.h.s., we would have found an homogeneous transformation, with the Lorentz transformation corresponding to a rotation around the $x^3$ axis. This second term is precisely what makes the polarization a non-covariant object and corresponds, note the $0$ index, to a boost, yielding the transformation group non-semi-simple. However, this is a no-loss situation for us, since requiring the non-homogeneous parts to cancel for any physical quantity will lead to constraints that will reduce the space of possible physical theories compatible with our assumptions.

In the case of higher spin particles, the corresponding transformation rules can be derived by considering the action of the operator $\Xi^\m_{\,\,\,\n}$ on the different pieces of the direct product.

\section{On-shell gauge invariance}
Let us now turn our attention to the interacting theory. In such a QFT where a set of massless particles interact in flat space-time, any individual process can be described by the corresponding element in the S-matrix. For a process with $n$ incoming and $m$ outgoing particles, this element will take the following schematic form in momentum space\footnote{This representation corresponds to a choice of normalization for the states in the Hilbert space of the form
\begin{align}
\langle q|p\rangle=2\,q^0(|q|)\, \delta^{(3)}(q-p)
\end{align}}
\begin{align}
S_{n\rightarrow m}=(\epsilon^{i_1}(q_{i_1}))^* (\epsilon^{i_2}(q_{i_2}))^*... (\epsilon^{i_m}(q_{i_m}))^*    \epsilon^{j_1}(q_{j_1}) \epsilon^{j_2}(q_{j_2})... \epsilon^{j_n}(q_{j_n})M_{i_1 i_2 ... i_m}^{j_1 j_2 ...j_n}
\end{align}
where the various $\epsilon^{k}(q_k)$ represent the polarizations of the different particles, with $i_k,j_k$ interpreted as sets of indices. These are contracted with the scattering amplitude $M_{i_1 i_2 ... i_m}^{j_1 j_2 ...j_n}$, which might be computed, in the perturbative regime, with the help of Feynman diagrams if we had a given lagrangian at our disposal. The scattering amplitude is thus a true tensor, transforming homogeneously under the action of the Lorentz group, and can only depend on the momentum and helicities of the particles involved in the scattering process.

Consider now a simple process of emission of a graviton of momentum $q^\m$ by the interaction of a set of fields. The S-matrix element of such a process will be just
\begin{align}\label{eq:S_emission}
S_{\pm}(q)=(\epsilon^{\m\n}_\pm(q))^* M_{\m\n}
\end{align}

This structure is pretty simple. All details of the interacting theory will be contained in the scattering amplitude $M_{\m\n}$, while contraction with the polarization selects only those terms which produce the physical graviton with light-like momentum.

In the following we will assume that the QFT producing such a S-matrix element is unitary. Thus, $S_{\pm}$ has to satisfy
\begin{align}
S_\pm S_\pm^\dagger=\mathbb{I}
\end{align}
in any physical frame. In particular, this implies that acting with a Lorentz transformation on $S_\pm(q)$ can, as most, modify it by adding a phase
\begin{align}\label{eq:S_transformation}
S_\pm(q)=e^{\pm 2i \bar{\theta}}S(\alpha q)
\end{align}
where $\bar{\theta}$ depends on the particular Lorentz transformation applied.

However, according to the particular representation \eqref{eq:S_emission}, when acting with a Lorentz transformation and using \eqref{eq:epsilon_graviton} and \eqref{eq:Lorentz_epsilon}, we find instead
\begin{align}
S_\pm(q)=e^{\pm 2 i \theta[\Xi]} \left([\alpha^{-1}]^\m_{\,\,\,\rho}-[\alpha^{-1}]^0_{\,\,\,\rho}\frac{q^\m}{|q|}\right)^* \left([\alpha^{-1}]^\n_{\,\,\,\sigma}-[\alpha^{-1}]^0_{\,\,\,\sigma}\frac{q^\n}{|q|}\right)^*\;(\epsilon^{\rho\sigma}_\pm(\alpha q))^* M_{\m\n}
\end{align}

For an infinitesimal Lorentz transformation $\alpha=\delta+\omega$ this reduces to
\begin{align}
S_\pm(q)= e^{\pm 2i \theta[\Xi]}S(\alpha q)-e^{\pm 2 i \theta[\Xi]}\frac{\omega^0_{\,\,\,\alpha}  }{|q|}\left[(\epsilon^{\mu\alpha}_\pm(q))^* q^\n+(\epsilon^{\nu\alpha}_\pm(q))^* q^\m\right]M_{\m\n}+O(\omega^2)
\end{align}

As we see, the first term precisely corresponds to the transformation \eqref{eq:S_transformation} imposed by unitarity. The second piece, coming from the boost part in the transformation \eqref{eq:Lorentz_epsilon} of the polarization, must thus drop out. Since the Lorentz transformation that we used is completely arbitrary, the only possible way to solve this discrepancy is to ask the scattering amplitude to satisfy
\begin{align}
(\epsilon^{\m\a}_\pm(q))^* q^\n M_{\m\n}=0
\end{align}
which enforces the condition
\begin{align}\label{eq:condition_graviton}
q^\m M_{\m\n}=\sigma q_\nu
\end{align}
where $\sigma$ is an undetermined proportionality factor.

What we have derived here is no other thing than the on-shell Ward identities for the scattering amplitude. These are required, when going from the lagrangian description to the S-matrix, to drop the non-physical polarizations out of any physical process. The word on-shell here, stating that this is satisfied only by physical states where $q^2=0$, is important, since off-shell and on-shell gauge invariance do not need to agree, as we will see later.

In order to understand this point, let us repeat the same construction for the S-matrix of emission of photons, whose polarization is $\epsilon^\m_\pm(q)$. The computation is identical to the one just showed for gravitons, one just needs to substitute the polarizations at all steps and take into account that the scattering amplitude now carries a single index $M_\m$. The on-shell Ward identity equivalent to \eqref{eq:condition_graviton} is
\begin{align}
q^\m M_\m=0
\end{align}
which agrees with the usual requirement of transversality for the scattering amplitude. This implies that any shift of the polarization of the form
\begin{align}\label{eq:U1}
\epsilon^\m_\pm(q)\rightarrow f(q)q^\m
\end{align}
with $f(q)$ an arbitrary function, will vanish and wont contribute to the S-matrix and, therefore, to any physical process. The polarization is not universally determined but instead there is a redundancy in choosing it corresponding to what we usually call a gauge symmetry. Actually, the transversality condition $q_\mu \epsilon^\mu_{\pm}=0$ can be regarded as a gauge choice for this symmetry, which is of course preserved by the physical states due to the fact that $q^2=0$, which makes the shift \eqref{eq:U1} irrelevant.

The situation is slightly more subtle for gravitons, though. In this case, the on-shell Ward identity \eqref{eq:condition_graviton} allows for a similar gauge redundancy as with the photon
\begin{align}\label{eq:on-shell_invariance_spin2}
\epsilon^{\m\n}_\pm(q)\rightarrow q^\m f^\n + q^\n f^\m
\end{align}
where, in order to preserve the gauge choice $q_\mu \epsilon^{\mu\nu}_{\pm}=0$, the vector $f^\mu$ must have only transverse components, satisfying $q_\mu f^\mu=0$.

If we were deriving this S-matrix from General Relativity in its weak field version, the Fierz-Pauli lagrangian of metric perturbations $h^{\m\n}$, we would have assumed that the gauge symmetry had to be the linearised diffeomorphisms group, corresponding to
\begin{align}
h^{\m\n}\rightarrow q^\m l^\n +q^\n l^\m
\end{align}
which looks similar to what we have found except for a little difference. The vector $l^\m$ is not constrained to be transverse but it also contains a longitudinal part. However, this longitudinal part can be always written as
\begin{align}
l^\m_L=q^\m L(q)
\end{align}
with $L(q)$ corresponding to a function of the momentum. This piece satisfies the constraint $q_\m l^\m_L=0$ as an identity due to the on-shell condition $q^2=0$ and thus it does not need to be imposed as a constraint on the scattering amplitude for what it concerns to the S-matrix. In other words, on-shell states are blind to the existence of this longitudinal piece and therefore its presence is not required for consistency of the on-shell symmetry. Indeed, it was shown in \cite{vanderBij:1981ym,Alvarez:2006uu} that the transverse transformations are enough to reduce the number of polarizations of the massless graviton down to two. 

The transverse transformations form a group that has been named \emph{TDiff} before \cite{Alvarez:2006uu} and they consist on the linearised version of volume preserving diffeomorphisms. Adding the longitudinal piece represents, when off-shell, the embedding of this in a bigger group, the linearised version of the whole diffeomorphism group. This means that, although the extra piece is invisible for on-shell dynamics, it will make a difference when defining off-shell fields, since their action needs to be invariant under the whole symmetry group that we choose to carry on.

This is one of the main points that we want to address here. Lorentz invariance of the S-matrix does not enforce the longitudinal part of the linearised diffeomorphim group to be a symmetry of physical states. Of course, it can always be taken to be so, as it is done in \cite{Weinberg:1964ew,Weinberg:1965rz}, but then this would lead eventually to the weak field limit of GR as the only possible lagrangian theory to describe the system. Here instead we will not impose the longitudinal part as a symmetry but we will solely stick to the transverse transformation imposed by \eqref{eq:condition_graviton}.
\section{The lagrangian description}
Let us now to try to derive the most general off-shell lagrangian which leads to the dynamics of gravitons described by the polarizations $\epsilon^{\m\n}_\pm(q)$ and enjoying the on-shell gauge invariance that we have just derived. In order to do this, let us introduce the concept of a vacuum state $|0\rangle$ and ladder operators $a^\dagger (q), a (q)$ which create and annihilate quanta of particles with momentum $q^\m$. The fields that give rise to gravitons are then given by
\begin{align}
\bar{A}^{\m\n}[x]=\sum_{\pm}\int \frac{d^3p}{(2\pi)^3} \;\epsilon^{\m\n}_{\pm}(p) \left(a(p)e^{ip\cdot x}+a^\dagger (p) e^{-ip\cdot x}\right)
\end{align} 
where we are summing over polarizations in order to create a parity even field\footnote{Alternatively we could construct a parity odd field by using $\epsilon^{\m\n}_+ - \epsilon^{\m\n}_-$. However, parity odd fields couple to anti-symmetric sources, which do not exist in flat-space. Thus, we will refrain to study those in what follows and we will stick to the even combination.}.

At this point, we can give a new interpretation to the on-shell gauge invariance \eqref{eq:on-shell_invariance_spin2}. Since the field is built in terms of the polarization, it means that it will not transform homogeneously under Lorentz transformations, but instead it will enjoy a transformation inherited from \eqref{eq:Lorentz_epsilon}. At an infinitesimal level $\alpha=\delta+\omega$, this transformation reads
\begin{align}
\epsilon^{\m\n}_{\pm}(q)= \epsilon^{\m\n}_{\pm}(\alpha q)-e^{\pm 2 i \theta[\Xi]}\frac{\omega^0_\alpha}{|q|}\left[(\epsilon^{\mu\alpha}_\pm(q))^* q^\n+(\epsilon^{\nu\alpha}_\pm(q))^* q^\m\right]+O(\omega^2)
\end{align}
which can be rewritten as
\begin{align}
\epsilon^{\m\n}_{\pm}(q)= \epsilon^{\m\n}_{\pm}(\alpha q)+q^\m f^\n + q^\n f^\m
\end{align}
with $f^\m$ satisfying $q_\m f^\m=0$, and given by
\begin{align}\label{eq:f}
f^\m=-e^{\pm 2 i \theta[\Xi]}\frac{\omega^0_\alpha}{|q|}(\epsilon^{\nu\alpha}_\pm(q))^*
\end{align}

We again find, although from a different point of view, the already known result that on-shell gauge invariance is precisely the redundancy required in order to cancel the non-physical degrees of freedom which obstruct Lorentz invariance of the theory.

From the transformation rule of the polarization, we derive the corresponding one for the field
\begin{align}
\bar{A}^{\m\n}[x]\rightarrow \bar{A}^{\m\n}[x]+\partial^\m \Phi^\n+\partial^\n \Phi^\m
\end{align}
where $\Phi^\m$ is the Fourier transform of $f^\m$ as given in \eqref{eq:f}, satisfying $\partial_\m \Phi^\m=0$.

At this point, let us note that the field here constructed satisfies $\bar{A}^{0\m}=\bar{A}^{\m 0}=0$, besides being traceless, symmetric, and transverse on its both indices. In what follows we will forget about full space-time indices and we will separate time, denoted by a zero, from space coordinates, which we will denote with latin letters. Therefore, we introduce the field\footnote{We will omit the $x$ dependence of the fields in position space for the rest of this work. It will be showed only when required for clarity of the discussion.} $\bar{A}^{ij}$ to describe the physical polarizations of the graviton, satisfying the following equations
\begin{align}
&\partial_i \bar{A}^{ij}=0\\
&\delta_{ij}\bar{A}^{ij}=0\\
&\square \bar{A}^{ij}=(\partial_i \partial^i-\partial_t^2)\bar{A}^{ij}=0
\end{align}
where the last line is just the statement that the field is on-shell, satisfying the Klein-Gordon equation and propagating massless degrees of freedom with a dispersion relation $(q^0)^2=|q|^2$, which was our starting assumption.

We wish to find which possible lagrangians allow for external gravitons of the kind we have described. For that, we will consider weak field lagrangians of at most second order in the fields and of the form
\begin{align}
{\cal L}_{phys}=-\frac{1}{2}\bar{A}^{ij}{\cal D}_{ijkl}\bar{A}^{kl}+\bar{A}^{ij}J_{ij}
\end{align}
with ${\cal D}_{ijkl}$ a differential operator and $J_{ij}$ being the source to which the graviton field couples.

If we were given this Lagrangian and insert it in a path integral, we could integrate out the fields by means of its propagator, which basically amounts to invert the operator ${\cal D}_{ijkl}$, and obtain a effective interaction between the sources of the form
\begin{align}
V[J]=\frac{1}{2}J_{ij}({\cal D}_{ijkl})^{-1}J_{kl}
\end{align}

Here we will do the opposite. Starting from what we know of gravitons we will compute the interaction and, by requiring it to be Lorentz covariant, we will find which constrains we must impose on the lagrangian description. Note that since we want to derive a fully covariant lagrangian, $J_{ij}$ must be the spatial part of some symmetric covariant source $J_{\m\n}$, which couples to the field through
\begin{align}
\int d^4x \bar{A}^{\m\n}J_{\m\n}
\end{align}

Moreover, since we do not want the non-physical gauge degrees of freedom to carry any interaction, we must require the coupling between the field and the source to be invariant under the on-shell transformation \eqref{eq:on-shell_invariance_spin2}
\begin{align}\label{eq:constraints_source}
\int d^4x \, \bar{A}^{\m\n}J_{\m\n}\rightarrow 2\int d^4x\, \partial^\m \Phi^\n J_{\m\n}=-2\int d^4x\, \Phi^\n \partial^\m J_{\m\n}
\end{align}
where we have used the symmetric character of the source and integrated by parts in the last step. Therefore, in order to cancel this coupling for arbitrary $\Phi$, we must require the source to satisfy
\begin{align}
\partial^\m J_{\m\n}=\partial_\n {\cal X}[x]
\end{align}
so that the integral vanishes after integration by parts, since $\partial_\m\Phi^\m=0$. Here ${\cal X}[x]$ is an in principle arbitrary function of the space-time coordinates.

It is important to note that gauge invariance here does not completely fix the form of the source, but instead allows for the existence of an arbitrary function. If we were repeating the same computation with a photon, or by assuming that the on-shell gauge invariance corresponds to the whole linearised diffeomorphisms group, as discussed in the last section; we would have found a different result, implying the whole conservation of the source and thus setting ${\cal X}[x]=0$. However, \emph{TDiff} invariance allows for this extra function and we will keep it arbitrary in what follows. 

Now we focus on computing the propagator for the field $\bar{A}^{ij}$. This is easy, since we know that $\bar{A}^{ij}$ is transverse and traceless, and it propagates massless degrees of freedom, thus its propagator must have a pole in $q^2=0$. The only option in momentum space is\footnote{We will define the propagator as
\begin{align}
{\cal D}_{ijkl}G^{klab}=-(\delta^{ia}\delta^{jb}+\delta^{ib}\delta^{ja})
\end{align}
} 
\begin{align}
G^{ijkl}(q)=-\frac{\Pi^{ijkl}}{q^2-i\epsilon}
\end{align}
where we are using the usual $i\epsilon$ prescription for the Feynman propagator.

The projector $\Pi^{ijkl}$ must project over those fields that are both symmetric in the pairs $\{i,j\}$ and $\{k,l\}$, as well as under $\{i,j\}\leftrightarrow \{k,l\}$; while at the same time being traceless and transverse in both pairs of indices. It can be constructed in the following way. We introduce the transverse projector $\Pi^{ij}$
\begin{align}
\Pi^{ij}=\delta^{ij}-\frac{q^i q^i }{|q|^2}
\end{align}
which ensures that we will select the transverse part of any element which we act upon with it. Thus, the following construction satisfies all the required conditions
\begin{align}
\Pi^{ijkl}=\frac{1}{2}\left(\Pi^{ik}\Pi^{jl}+\Pi^{il}\Pi^{jk}-\Pi^{ij}\Pi^{kl}\right)
\end{align}

Now, this is not covariant, in particular because of the presence of only space components, so it wont lead to a covariant interaction when contracted with the sources. In order to guess what is happening here, let us use the trick of \cite{Weinberg:1965rz} and introduce a time-like vector $n^\m=(1,0,0,0)$ so that we can rewrite the spatial momentum $q^i$ in terms of the full four-momentum $q^\m$
\begin{align}
(0,q^i)\equiv q^\m -n^\m q^0
\end{align}

The transverse projector can equally be rewritten as
\begin{align}
\Pi^{\m\n}=\begin{pmatrix}
0&0\\
0& \Pi^{ij}
\end{pmatrix}\equiv g^{\m\n}+\frac{n^\m q^\n +n^\n q^\m}{|q|^2}q^0 +\frac{n^\m n^\n}{|q|^2}q^2 -\frac{q^\m q^\n}{|q|^2}
\end{align}
and therefore the propagator can be cast into the form
\begin{align}
G^{ijkl}(q)\equiv G^{\m\n\a\b}(q)=-\frac{1}{2}\left(\Pi^{\m\a}\Pi^{\n\b}+\Pi^{\m\b}\Pi^{\n\a}-\Pi^{\m\n}\Pi^{\a\b}\right)\frac{1}{q^2-i\epsilon}
\end{align}
which, let us stress, is not a covariant object due to the presence of the time-like vector $n^\mu$. The advantage of this form is that we can now compute the effective action for $J_{\m\n}$ easily as
\begin{align}
V[J]=\frac{1}{2}J_{\m\n}\frac{\Pi^{\m\n\a\b}}{q^2-i\epsilon}J_{\a\b}
\end{align}
which splits in three parts
\begin{align}\label{eq:interaction}
\nonumber V[J]&=\frac{1}{2}\frac{1}{q^2-i\epsilon}\left(J_{\m\n}J^{\m\n}-\frac{1}{2}J^2 +2{\cal X}J-4{\cal X}^2\right)\\
\nonumber &+\frac{1}{2 |q|^2}\left( 2 J_\m^{\,\,\,0}J_0^{\,\,\,\m}+2{\cal X}^2 -J{\cal X} -J J_{00}\right)\\
&-\frac{q^2}{2|q|^4}\left(\frac{1}{2}(J_{00})^2+{\cal X}J_{00}+\frac{1}{2} {\cal X}^2\right) 
\end{align}
where $J=\eta^{\m\n}J_{\m\n}$ and we have used the source conservation constraint $\partial_\mu J^{\m\n}=\partial^\nu {\cal X}$.

Before going further, let us note that the interaction \eqref{eq:interaction} can be rewritten in a simpler form
\begin{align}
V[J]=\frac{1}{2}\frac{1}{q^2-i\epsilon}\left(\tilde{T}_{\m\n}\tilde{T}^{\m\n} -\frac{1}{2} T^2  \right)+\frac{1}{2|q|^2}\left( 2 \T_{i0}\T^{i}_0 -\T_{00}\T_{00}-\T_{ii}\T_{00}\right)-\frac{q^2}{4|q|^4}\T_{00}\T_{00}
\end{align}
where $\T_{\m\n}=J_{\m\n}-{\cal X}\eta_{\m\n}$.

This new source $\T_{\m\n}$ is not only symmetric, but also conserved, due to the fact that $\partial^\m J_{\m\n}=\partial_\n {\cal X}$. Since the only conserved symmetric source available in flat space-time is the energy-momentum tensor $T_{\m\n}$, this must differ from it as most in a constant piece, which can be always absorbed in the arbitrary function ${\cal X}$. Thus, we can identify $\T_{\m\n}$ and $T_{\m\n}$, having
\begin{align}
J_{\m\n}=T_{\m\n}+{\cal X}\eta_{\m\n}
\end{align}

This intermediate result, that ${\cal X}$ drops out from the interaction when written in this way, was to be expected, since it is no more than the statement of the \emph{weak equivalence principle} in the weak field approximation -- gravitons interact only with the conserved energy-momentum tensor, regardless of what other details or technical tools we are introducing in the theory. What we have done here is deriving the weak version of the equivalence principle from the propagation of massless gravitons. We leave out the question of the strong version of the principle, though. For that, we would need to introduce a self-consistent interaction between gravitons which preserves all the hypothesis previously discussed. Although it seem plausible, we will not get into that matter here.

Now, let us come back to the interaction \eqref{eq:interaction}. The first element in there represents indeed a covariant interaction propagated by a massless particle, thus the pole in $q^2=0$. The other two, however, are more problematic. First of all, they are not covariant, and even worse, they contain poles in $|q|^2$, so they will lead to instantaneous interactions and  non-local contributions when taken to position space by a Fourier transform. Clearly, if we were to formulate a theory containing only the variables that we have carried all along until here, we would be in danger, unable to find a covariant interaction. 

The problem can be solved by appending to $\bar{A}^{ij}$ a new set of auxiliary variables  in such a way that their contribution to the interaction will cancel the dangerous terms in \eqref{eq:interaction}. It is important to stress, though, that we are not adding new degrees of freedom by doing that, because their propagators, producing the last two lines of \eqref{eq:interaction} with opposite sign, will not have poles in $q^2$ and therefore they will not propagate physical excitations\footnote{It is instructive to take a look to the situation for photons. In that case, the interaction will be between a transverse field $A^i$ and a source $j^i$, and takes the form 
\begin{align}
V[j]=\frac{j_\m j^\m}{q^2-i\epsilon}+\frac{(j_0)^2}{|q|^2}
\end{align}
The requirement to cancel the non-covariant piece leads to the introduction of a fourth component $A^0$ whose equation of motion is the Coulomb's law $\partial_i \partial^i A^0=-j_0$.}.

We thus introduce three new auxiliary variables, two scalars $A^{00}$ and $A^{ii}$, and a vector $A^{i0}$, which must couple to $T_{00}$, $T_{ii}$ and $T_{i0}$ respectively through a Lagrangian
\begin{align}\label{eq:L_aux}
{\cal L}_{aux}=-\frac{1}{2}(A^{00},A^{ii},A^{i0})\mathfrak{M}\begin{pmatrix}
A^{00}\\
A^{jj}\\
A^{i0}
\end{pmatrix}+\left( A^{00}T_{00}+A^{i0}T_{i0}+A^{ii}T_{jj}\right)
\end{align}
where $\mathfrak{M}$ is a matrix valued differential operator. In principle we could choose a different combination of couplings in the scalar sector, for instance
\begin{align}\label{eq:invertible_trans}
(A^{00},A^{ii})\begin{pmatrix}
a&b\\c&d
\end{pmatrix} \begin{pmatrix}
T_{00}\\ T_{ii}
\end{pmatrix}
\end{align}
with $a,b,c$ and $d$ constants. However, we can always rotate the fields $A^{00}$ and $A^{ii}$ to diagonalize this interaction, ending up with \eqref{eq:L_aux} after a rescaling. Note that for all these rotations to be equivalent, we must demand the matrix in the middle to be diagonalizable and thus invertible. Singular transformations will play an important role later, but for the moment we will consider only invertible ones.

The value of $\mathfrak{M}$ must be chosen carefully in order to cancel the non-covariant contributions in \eqref{eq:interaction}. Thus, we have in momentum space
\begin{align}
\mathfrak{M}^{-1}=\frac{1}{|q|^2}\begin{pmatrix}
-\frac{q^2}{|q|^2}-2& -1&0\\
-1&0&0\\
0&0&4
\end{pmatrix}
\end{align}

Inverting this matrix we have
\begin{align}
\mathfrak{M}=\begin{pmatrix}
0&-|q|^2&0\\
-|q|^2& -q^2-2|q|^2&0\\
0&0&\frac{|q|^2}{4}
\end{pmatrix}
\end{align}
so the lagrangian in position space is
\begin{align}
{\cal L}_{aux}=\frac{1}{4} A^{i0} \Delta A^{i0}- 2 A^{ii}\Delta A^{jj}+A^{ii}\square A^{jj}-2A^{ii}\Delta A^{00}+\left( A^{00}T_{00}+A^{i0}T_{i0}+A^{ii}T_{ii}\right)
\end{align}

To this, we need to append the Lagrangian for the physical modes $\bar{A}^{ij}$. This is simply
\begin{align}
{\cal L}_{phys}=-\frac{1}{2}\bar{A}^{ij}\square\bar{A}_{ij}+\bar{A}^{ij}T_{ij}+ \lambda\left(\partial_i \bar{A}^{ij}\partial_k \bar{A}^{k}_j\right)
\end{align}
where $\lambda$ is a Lagrange multiplier that cancels the longitudinal components out in the limit $\lambda\rightarrow\infty$, ensuring transversality of the field.

The whole lagrangian ${\cal L}_{phys}+{\cal L}_{aux}$ now satisfies all the requirements that we want. It propagates physical massless gravitons and it leads to a covariant interaction in the effective action for the source. However, the interaction piece does not look covariant at first sight. In order to solve this, let us rescale the auxiliary fields in the following way
\begin{align}
A^{i0}\rightarrow -2A^{i0}\qquad A^{ii}\rightarrow \frac{1}{3}A^{ii}
\end{align}

Since this is just a global rescaling, it will not lead to any new dynamics, but it allows us to rewrite the lagrangian as
\begin{align}
{\cal L}&= A^{i0} \Delta A^{i0}+ \frac{2}{9} A^{ii}\Delta A^{jj}+\frac{1}{9} A^{ii}\square A^{jj}-\frac{2}{3} A^{ii}\Delta A^{00}-\frac{1}{2}\bar{A}^{ij}\square\bar{A}_{ij}+\lambda(\partial_i \bar{A}^{ij}\partial_k\bar{A}^{k}_j)+A^{\m\n}T_{\m\n}
\end{align}
with a now covariant interaction term, after defining the field
\begin{align}
A^{\m\n}=\begin{pmatrix}
A^{00}&A^{i0}\\
A^{i0}&A^{ij}
\end{pmatrix}
\end{align}
where $A^{ij}=\bar{A}^{ij}+\frac{1}{3}A^{kk}\delta^{ij}$.

However, the real source of the theory is not $T_{\m\n}$ but $J_{\m\n}=T_{\m\n}+\eta_{\m\n}{\cal X}$. For the physical field there is no difference, since we have seen that all dependence in ${\cal X}$ cancels out. However, it makes a difference for the auxiliary fields in the scalar sector, modifying \eqref{eq:L_aux} to be
\begin{align}
{\cal L}_{aux}=-\frac{1}{2}(A^{00},A^{ii},A^{i0})\mathfrak{M}\begin{pmatrix}
A^{00}\\
A^{jj}\\
A^{i0}
\end{pmatrix}+\left( A^{00}T_{00}-2A^{i0}T_{i0}+\frac{1}{3}A^{ii}T_{jj}+ (A_{ii}-A_{00}){\cal X}\right)
\end{align}
where we have already rescaled the fields. This will induce an extra contribution in the effective action of the form
\begin{align}\label{eq:V_extra}
V_{\text{extra}}=\frac{{\cal X} \left(q^2 (2 T_{00}-{\cal X})+2 |q|^2 (-T_{00}+T_{ii}+2 {\cal X})\right)}{4
   |q|^4}
\end{align}
where the dependence on ${\cal X}$ has not cancelled.

The most general lagrangian which propagates spin two degrees of freedom in a Lorentz invariant way is thus
\begin{align}\label{eq:final_lagrangian}
{\cal L}&= A^{i0} \Delta A^{i0}+ \frac{2}{9} A^{ii}\Delta A^{jj}+\frac{1}{9} A^{ii}\square A^{jj}-\frac{2}{3} A^{ii}\Delta A^{00}-\frac{1}{2}\bar{A}^{ij}\square\bar{A}_{ij}+\lambda(\partial_i \bar{A}^{ij}\partial_k\bar{A}^{k}_j)+A^{\m\n}J_{\m\n}
\end{align}
and leads to the interaction we are looking for but it will also induce the extra piece \eqref{eq:V_extra} which cannot be cancelled if we keep the arbitrary character of the source.


\section{Off-shell gauge invariance}
The lagrangian \eqref{eq:final_lagrangian} leads to the theory of gravitons that we were looking for and couples to the right source, defined by on-shell gauge invariance in \eqref{eq:on-shell_invariance_spin2}. However, the construction is still not complete, since there is an explicit presence of an undetermined function ${\cal X}$ which represents a new source and which induces a new contribution to the effective potential depending on the value of ${\cal X}$.

Additionally, there is the question of the origin of ${\cal X}$. It is a source for the gravitational field, so it must be constructed out of matter fields. However, the fact that it does not gravitate in the standard sense, the effective action induced by physical degrees of freedom containing only the energy-momentum tensor in order to comply with the equivalence principle, tangles up any effort of giving an explicit definition. Since we want our theory to be precisely defined, we need to eliminate the contribution of ${\cal X}$ from the effective potential, erasing the ambiguity.

Looking to \eqref{eq:V_extra}, we see that the simplest way is to enforce ${\cal X}=0$, so that we directly identify the source with the energy-momentum tensor
\begin{align}
J_{\m\n}=T_{\m\n}
\end{align}

In this case, the source is now fully conserved $\partial^\m J_{\m\n}=0$ and, reverting the argument in \eqref{eq:constraints_source}, we see that the field must be now invariant under the whole set of linearised diffeomorphisms in order to avoid the contribution of gauge modes to the effective action
\begin{align}
A^{\m\n}\rightarrow A^{\m\n}+\partial^\m f^\n +\partial^\n f^\m
\end{align}
where now $f^\m$ is unconstrained, the transformation containing both transverse and longitudinal diffeomorphisms when evaluated off the mass-shell. Importantly, the longitudinal part of the transformation acts here by protecting the value  ${\cal X}=0$, ensuring stability of the action under renormalization in the absence of anomalies.

If we do so, the reduced lagrangian obtained from \eqref{eq:final_lagrangian} can be derived from the \emph{Fierz-Pauli} theory, the weak field limit of GR, for a symmetric tensor $h^{\m\n}$
\begin{align}\label{eq:Fierz-Pauli}
{\cal L}_{FP}=-\frac{1}{4}h^{\m\n}\square h_{\m\n}+\frac{1}{4}h\square h-\frac{1}{2}\partial_\m h \partial_\n h^{\m\n}+\frac{1}{2}\partial_\m h^{\m\n}\partial_\alpha h^\alpha_\n
\end{align}

If we fix the gauge symmetry of this theory, corresponding to linearised \emph{Diff} and thus requiring of four conditions, by imposing
\begin{align}\label{eq:gauge_conditions}
\partial_i \left(h^{ij}-\frac{1}{3}\delta^{ij} h^{kk}\right)=0,\quad \partial_ih^{i0}+\frac{2}{3}\partial_0 h^{kk}=0
\end{align}
we recover the lagrangian \eqref{eq:final_lagrangian}, up to the use of the equations of motion and total derivatives. Thus, S-matrix unitarity and Lorentz invariance eventually lead to the Lagrangian of a spin two field that we have known for decades.

Since any combination of the auxiliary scalar fields $A_{00}$ and $A_{ii}$ can be taken to the diagonal form that we have used so far, it seems that this is the only option to get a theory of gravitons -- to enforce ${\cal X}=0$ and increase the symmetry to full linearised diffeomorphisms. However, this is only true for the case in which we can diagonalize the matrix in \eqref{eq:invertible_trans}. In principle, we could also consider rotations of the fields which lead to singular matrices
\begin{align}\label{eq:singular_rotation}
\tilde{A}_{00}=\alpha A_{00}+\beta A_{ii},\qquad 
\tilde{A}_{ii}=\gamma\left(  \alpha A_{00}+\beta A_{ii} \right)
\end{align}

These cannot be inverted and thus represent an isolated set of theories unrelated to the invertible rotations. For these transformations, the coupling to ${\cal X}$ becomes
\begin{align}
(\gamma -1)(A_{00}-A_{ii}){\cal X}
\end{align}
which vanishes only when $\gamma=1$, implying $\tilde{A}_{00}=\tilde{A}_{ii}$ or, in other words, that the new $\tilde{A}_{\m\n}$ must be traceless. That is, there exist a singular field redefinition, corresponding to defining a theory in which the trace is not dynamical, that  solves the problem as well. 

This seems overkilling and it is indeed subtle, since the field rotation is not invertible and seems to reduce the number of independent variables in the action. However, the out-coming theory is perfectly regular. Since we can always adjust the value of ${\cal X}$ to absorb any difference in the trace without modifying the value of the on-shell effective action, we could always have avoided to introduce the singular rotation and instead have chosen to define ${\cal X}$ it as
\begin{align}
{\cal X}=-\frac{1}{4}J+{\cal Y}
\end{align}
where ${\cal Y}$ is again arbitrary. By doing this, the interaction part of the lagrangian takes the form
\begin{align}
\nonumber {\cal L}_{int}&=\left(J_{\m\n}-\frac{1}{4}J \eta_{\m\n}\right)A^{\m\n}+(A^{00}-A^{ii}){\cal Y}=\left(A_{\m\n}-\frac{1}{4}A\eta_{\m\n}\right)J^{\m\n}+(A^{00}-A^{ii}){\cal Y}=\\
&=\left(A_{\m\n}-\frac{1}{4}A\eta_{\m\n}\right)T^{\m\n}+(A^{00}-A^{ii}){\cal Y}
\end{align}
where we have used the fact that $J^{\m\n}=T^{\m\n}+{\cal X}\eta^{\m\n}$.

In this way we isolate all the arbitrariness in the definition of ${\cal X}$ into the interaction with the trace of the graviton field, while the physical source $J_{\m\n}$ only couples to the traceless part of $A^{\m\n}$. This again shows that it is enough to define a theory for this traceless part in order to recover the interaction \eqref{eq:interaction} and at the same time get rid of the problems associated to the arbitrariness of ${\cal X}$, forgetting about introducing any interaction with the trace, which is dynamically irrelevant at this level. Being this equivalent to performing the singular transformation \eqref{eq:singular_rotation}, the latter must lead to a perfectly regular theory.

Let us then introduce the change of variables that makes $A^{\m\n}$ explicitly traceless, corresponding to the covariant realization of the singular field rotation \eqref{eq:singular_rotation}
\begin{align}\label{eq:change_traceless}
A^{\m\n}\rightarrow \tilde{A}^{\m\n}=A^{\m\n}-\frac{1}{4}\eta^{\m\n}\eta_{\a\b}A^{\a\b}
\end{align}

The result, which can be read from \eqref{eq:final_lagrangian} after elementary algebra, also corresponds to a known lagrangian. Incidentally, it corresponds to the gauge fixed version of the only other known Lorentz invariant lagrangian which propagates a single spin-two physical degree of freedom, the \emph{WTDiff} lagrangian of \cite{Alvarez:2006uu}
\begin{align}\label{eq:WTDiff}
{\cal L}_{WTDiff}=-\frac{1}{4}h^{\m\n}\square h_{\m\n}+\frac{3}{32}h\square h-\frac{1}{4}\partial_\m h \partial_\n h^{\m\n}+\frac{1}{2}\partial_\m h^{\m\n}\partial_\alpha h^\alpha_\n+T_{\m\n}h^{\m\n}
\end{align}

This is a theory that, besides being invariant under \emph{TDiff}, also enjoys a linearised Weyl invariance $h_{\m\n}\rightarrow h_{\m\n}+\phi \eta_{\m\n}$ with $\phi$ an arbitrary function, which ensures the irrelevance of the trace and thus the \emph{W} in its name. By taking the \emph{WTDiff} lagrangian and fixing the gauge by using the traceless version of \eqref{eq:gauge_conditions}, one recovers exactly the traceless version of \eqref{eq:interaction}. Again, the increased symmetry, in this case the Weyl part of the linearised gauge group, protects the dynamics of the theory not by forbidding ${\cal X}$ this time but by ensuring its irrelevance at the dynamical level.

The \emph{WTDiff} theory, besides being completely equivalent at the dynamical level to the Fierz-Pauli theory \cite{Alvarez:2012px,Alvarez:2005iy}, also keeps the arbitrary character of the source in its core as a consequence of Weyl invariance. Any shift of the form $T_{\m\n}\rightarrow T_{\m\n}+\eta_{\m\n}{\cal X}$ is non-physical. As such, it does not source the equations of motion and thus vacuum solutions, defined as solutions to the equations of motion when the sources vanish, are still solutions even for non-vanishing ${\cal X}$. This is contrary to what happens in General Relativity, or the linearised Fierz-Pauli theory, where ${\cal X}$ sources the trace of the (linearised) Einstein equations and modifies the solution. Of course, this comes at the cost of having a more complicated gauge structure, leading to an open algebra in the ghost sector and requiring more sophisticated techniques for its quantization\cite{Alvarez:2015sba}, but maybe it is not such a big prize to pay.

Finally, it is worth to comment and emphasize that the two options here discussed, the Fierz-Pauli and \emph{WTDiff} lagrangians, are the only possible theories that describe dynamical gravitons and that can be found with this method\footnote{We allow for the possible existence of some exotic theory that might violate our assumptions here, although it seems difficult for such a theory to exist as a local QFT.}, corresponding to invertible transformations and to the unique singular rotation that solves the issue. Any other option will lead to the extra interaction \eqref{eq:V_extra} which cannot be cancelled unless we set ${\cal X}=0$ or we perform the traceless transformation \eqref{eq:singular_rotation} with $\gamma=1$.

\section{The non-linear theories}
The two theories that we have derived in the previous section correspond both to the weak field limit of known gravitational theories. In the case of the Fierz-Pauli lagrangian, it is known that it reproduces the lowest order in the weak field limit of General Relativity. Indeed, let us take the Einstein-Hilbert action
\begin{align}\label{eq:Einstein-Hilbert}
S=M_p^2 \int d^4x \sqrt{|g|}\,R
\end{align}
with $M_p$ being the Planck mass. The weak field theory is defined by expanding the metric around flat space-time
\begin{align}\label{eq:expansion}
g_{\m\n}=\eta_{\m\n}+\frac{1}{M_p}h_{\m\n}
\end{align}
where the factor of $M_p$ is included in order to get a canonical kinetic term for $h_{\m\n}$. After plugging this expansion into \eqref{eq:Einstein-Hilbert}, the second order action exactly corresponds to the Fierz-Pauli lagrangian \eqref{eq:Fierz-Pauli}, while higher orders will contain interactions leading to reproducing the full dynamics of GR.

This derivation goes in the direction from GR to Fierz-Pauli. There is a way to perform the converse, though, introduced in \cite{Deser:2009fq}\footnote{There are some concerns, however, with the structure of certain terms in the expansion of the graviton, since they  seem to be non-analytic in the coupling constant. See discussion in \cite{Padmanabhan:2004xk,Butcher:2009ta}.}. Starting from the Fierz Pauli lagrangian and by introducing consistent self-interactions, the author in \cite{Deser:2009fq} is able to re-sum the series, eventually finding the Einstein-Hilbert action.

For years, it was thought that this path was unique. That starting from graviton self-interactions, the only possible non-linear theory to be found was GR. However, this claim was shown wrong in \cite{Barcelo:2014mua}, where the authors show  that the\emph{WTDiff} theory is equally consistent and can be completed to the action of Unimodular Gravity
\begin{align}
S_{UG}=M_p^2 \int d^4x \, R_{|g|=1}
\end{align}
which is written as the Einstein-Hilbert action constrained to metrics of unit determinant. For computational purposes, it can be recast to an unconstrained form where we make the unimodular condition explicit by a change of variables\cite{Alvarez:2012uz} 
\begin{align}
g_{\m\n}\rightarrow |g|^{-\frac{1}{4}}g_{\m\n}
\end{align}
so that we have
\begin{align}\label{eq:action_unimodular}
S_{UG}=M_p^2\int d^4x |g|^{\frac{1}{4}}\left(R+\frac{3}{32}\frac{\nabla_\m |g|\nabla^\m |g|}{|g|^2}\right)
\end{align}
where none of the variables here is constrained any more. Remarkably, this change of variable corresponds, for the weak field approximation, to the traceless transformation \eqref{eq:change_traceless}.

In this form, the action of Unimodular Gravity is invariant under volume preserving diffeomorphisms\footnote{That is, those diffeomorphims that preserve the value of $\sqrt{|g|}$.} as well as under Weyl transformations
\begin{align}
g_{\m\n}\rightarrow \Omega^{2}(x) g_{\m\n}
\end{align}

The linear version of these symmetries are precisely the gauge symmetries of the \emph{WTDiff} theory and its full lagrangian can be obtained from \eqref{eq:action_unimodular} by performing a weak field expansion like in \eqref{eq:expansion}.

As we discussed, the advantage of the \emph{WTDiff} theory over Fierz-Pauli is the fact that, thanks to the arbitrariness of the function ${\cal X}$ in the definition of the source, a possible cosmological constant cannot have an effect on the dynamics and does not prevent flat space-time to be a solution of the equations of motion around which to define a sensible perturbative QFT. This property emerges in the action $\eqref{eq:action_unimodular}$ because Weyl invariance forbids the coupling $\sqrt{|g|}\Lambda$ at the level of the lagrangian. Moreover, because of Weyl invariance and trace irrelevance, UG only couples to the \emph{traceless} part of $T_{\m\n}$, so that the presence of vacuum energy sourced by matter fields is also irrelevant.

Although this looks overkilling, since it seems to forbid the dynamics driven by vacuum energy and thus any effect that could lead to the observed accelerated expansion of the Universe, this is not really true. By deriving the equations of motion of the action \eqref{eq:action_unimodular} in the gauge $|g|=1$ we find the traceless part of the Einstein equations
\begin{align}\label{eq:traceless_Einstein_equations}
R_{\m\n}-\frac{1}{4}g_{\m\n}R=\frac{1}{M_p^2}(T_{\m\n}-\frac{1}{4}g_{\m\n}T)
\end{align}

To this, we must append the fact that the second Bianchi identities must be satisfied not only as a dynamical consequence of \emph{TDiff} invariance but also as a geometrical constraint for the space-time manifold
\begin{align}\label{eq:bianchi}
\nabla_\m R^{\m\n}=\frac{1}{2}\nabla^\n R
\end{align}

Taking the derivative of \eqref{eq:traceless_Einstein_equations} and using \eqref{eq:bianchi} and the conservation of the energy-momentum tensor, we find the following constraint
\begin{align}
\frac{1}{4}\nabla_\m \left(R+\frac{1}{M_p^2}T\right)=0
\end{align}
which implies
\begin{align}
R+\frac{1}{M_p^2}T=4\lambda
\end{align}
where $\lambda$ is an \emph{integration constant}.

Plugging this back on the traceless equations \eqref{eq:traceless_Einstein_equations} we recover the whole set of Einstein equations
\begin{align}
R_{\m\n}-\frac{1}{2}g_{\m\n}+\lambda g_{\m\n}=\frac{1}{M_p^2}T_{\m\n}
\end{align}
with the integration constant $\lambda$ taking the role of a cosmological constant. However, unlike $\Lambda$ or vacuum energy, its value is stable under radiative corrections and it therefore does not suffer from this cosmological constant problem. Its value, whatever it is, must be fixed solely by the given initial conditions when solving the equations of motion or, in the QFT language, by an experiment measuring it, after which the result is not renormalized.

\section{Discussion and conclusions}
Through this work we have studied how much a lagrangian formulation of a theory of gravitons is constrained by basic properties of the S-matrix. By assuming unitarity and Lorentz invariance of physical quantities, we have seen that the lagrangian describing massless gravitons is not unique but instead we found two possibilities, corresponding to the well-known Fierz-Pauli theory and to the so-called \emph{WTDiff} theory. The key point to understand this duality is that the on-shell gauge invariance for the physical states is the same for both theories, the transverse part of the linearised diffeomorphism group, which we have denoted by \emph{TDiff}. We have also seen that, regardless of the lagrangian, we need to append to the physical degrees of freedom a set of auxiliary ones, which do not lead to the propagation of particles but contribute to the effective action of the sources with Coulomb-like interactions needed to cancel non-covariant pieces.

In the case of Fierz-Pauli, the symmetry group is enhanced by the presence of a longitudinal transformation, which is tautological on-shell but off-shell leads to the conservation of the current $J_{\m\n}$ of the theory, which becomes then the energy-momentum tensor $T_{\m\n}$. The other possible completion is provided in the \emph{WTDiff} theory, where we drop the interaction with the trace of the graviton field while nevertheless finding the same effective action for the source, which is now the traceless part of $T^{\m\n}$ and still satisfies the equivalence principle. Both theories enjoy non-linear completions by following the Deser trick. In the Fierz-Pauli case, General Relativity emerges from self-interactions, while Unimodular Gravity is obtained from the \emph{WTDiff} lagrangian.

At this point it is thus natural to ask ourselves if there is a reason to prefer one theory over the other, apart from historical reasons. As far as we know, both GR and UG are dynamically equivalent both in the weak field limit and classically, and they share the same S-matrix in flat space. Moreover, they are the only known Lorentz invariant theories that propagate only a massless spin two degree of freedom and also the only ones that can be completed from the weak field approximation to the non-linear level. 

However, both theories are not identical. In particular, the problem of the radiative corrections to the cosmological constant appears in a different light in Unimodular Gravity. By not coupling to the trace of $T_{\m\n}$, vacuum energy does not gravitate at all and the problem is avoided. The cosmological constant is not a coupling in the lagrangian but an integration constant whose value must be set by initial conditions. It is true that there is left the problem of fixing this and to explain why it does take such an unnatural value on our Universe but at least, once the cosmological constant is fixed, radiative corrections do not modify its value producing a hierarchy problem.

The other difference that can be pointed out at this level is the fact that symmetry groups are different for GR and UG. While the former enjoys the full diffeomorphisms as a gauge symmetry, the latter is invariant only under those that preserve the volume, plus Weyl transformations. This leads to a problem in identifying physically equivalent solutions to the equations of motion, since the gauge orbits are different. In particular, this could lead to radical differences in the interpretation of horizons and the interior of black holes, where certain changes of frames are needed in order to avoid apparent singularities in the metric. 

Last but not least, it has been pointed out by Padmanabhan\cite{Padmanabhan:2004xk} that the QFT formulation of Unimodular Gravity could be actually advantageous over that of General Relativity, since it can be written exactly as resembling a Yang-Mills theory, while this is not completely possible in GR. In any case, as far as we know, UG is as good as GR for many matters and leads to the same physics for as far we have tested gravity in our Universe. It would be good to find a way to discriminate between the two theories.

\section*{Acknowledgements}
I am grateful for the hospitality of the Instituto de Física Teórica UAM-CSIC, where part of this work was done. I also wish to thank Enrique Álvarez and Carmelo P. Martin for fruitful discussions and/or email exchange and to Sergey Sibiryakov for discussions and very useful comments on a previous version of this text. My work has been supported by the Tomalla Foundation.

\bibliography{smatrix}{}
\bibliographystyle{hunsrt}
\end{document}